# Hubbard Model with Lüscher fermions – a progress report

P. Sawicki[a] and J. Wosiek[b] [*] [†]

[a]Institute of Physics, Jagellonian University,
Reymonta 4, Cracow, Poland

[b]Institute of Computer Science, Jagellonian University,
Reymonta 4, Cracow, Poland

Some modifications of the Lüscher algorithm, which reduce the autocorrelation time, are proposed and tested.

## 1. HUBBARD MODEL IN THE LÜSCHER REPRESENTATION

The problem of dynamical fermions still attracts new interest. Although solved in principle still provides the considerable challenge in practice. To remind, the main difficulty with simulating the theory with dynamical fermions consists of the nonlocality of the fermionic determinant

$$\int [d\Psi d\overline{\Psi}] \exp(-\overline{\Psi}\mathcal{M}\Psi) \sim det(\mathcal{M}), \quad (1)$$

which depends functionally on the gauge field in the case of QCD for example. The numerical complexity of best known algorithms, namely hybrid Monte Carlo, scale with the volume of the system as $V^{4/3}$ [1]. However these algorithms suffer from the strong correlations between generated configurations. Therefore the Lüscher's proposal of algorithm with numerical complexity $V$ is very promising [2, 3].

In this report we present some new results of applying the Lüscher method to the Hubbard model. Our aim at present stage is to better understand strong correlations, existing also in this approach [4], and possibly to offer some remedy of this problem. Our findings are rather general and can be relevant also for other systems.

The Hamiltonian of the Hubbard system reads

$$H = -K \sum_{<ij>\sigma} a_{i\sigma}^\dagger a_{j\sigma} - \frac{U}{2} \sum_i (n_{i\uparrow} - n_{i\downarrow})^2 \quad (2)$$

$$+ \mu \sum_{i\sigma} n_{i\sigma}, \quad (3)$$

[*]Presented by P. Sawicki.
[†]Supported in part by the KBN grants no PB P03B19609 and PB 2P30225206.

where $n_{i\sigma} = a_{i\sigma}^\dagger a_{i\sigma}$, and $a_{i\sigma}$ denotes the creation operator of an electron at the lattice site $i$ and with the spin $\sigma = \uparrow, \downarrow$. The physical parameters entering the model are: K - hopping parameter, U - the strength of the effective Coulomb interaction, $\mu$ - the chemical potential and the inverse temperature $\beta$. The detailed derivation of the Lüscher representation of the Hubbard model can be found elsewhere [5]. The final form of the partition function reads

$$Z \simeq \int [dAd\phi] \exp\left(-\int A^2(x)/2 \, d^3x\right) \quad (4)$$

$$\exp\left(-\int \sum_{k=1}^N \phi_k^\dagger \left[(\mathcal{Q}^\dagger\mathcal{Q} - \alpha_k)^2 + \beta_k^2\right] \phi_k d^3x\right) \quad (5)$$

Where $\mathcal{Q}^\dagger\mathcal{Q} \equiv \mathcal{M}^\dagger\mathcal{M}/\lambda_{max}$ and $\lambda_{max}$ denotes the largest eigenvalue of $\mathcal{M}^\dagger\mathcal{M}$, $\mathcal{M}$ being the original fermion matrix entering the Euclidean formulation.

$$\Psi^*\mathcal{M}\Psi = \frac{K\beta}{N_t} \sum_{<ij>t} \Psi_{it}^* \Psi_{jt} + \sum_{it} \Psi_{it}^*(\Psi_{it} - \Psi_{it-1})$$

$$+ \sum_{it} \Psi_{it}^* \Psi_{it} \left(\exp\left[\sqrt{\frac{U\beta}{N_t}} A_{it} - U\frac{\beta}{N_t}\right] - 1\right).$$

$A(x)$ is the Hubbard-Stratanovich continuous field, and $\phi_k(x)$ are the auxiliary bosonic fields introduced by Lüscher, $\alpha_k$ and $\beta_k$ being the real and imaginary parts of the zeros of the polynomial $P_N(x)$ approximating the inverse $P_N(x) \simeq 1/x$.

The Lüscher approximation is uniform in the interval $(\epsilon, 1)$ and the error falls exponentially with the number of bosonic fields. The value of $\epsilon$ is determined by the width of $\mathcal{Q}^\dagger\mathcal{Q}$ spectrum. We have kept the relative error at the level of $O(10^{-4})$. This required $100 - 150$ bosonic fields in the simulations.

## 2. RESULTS AND DISCUSSION

The runs have been done mainly on the $6^2 8$ lattice. One MC step consists of the single heat bath generation for $\phi$ fields and ten Metropolis updates of the Hubbard-Stratonovich field $A$. Additionally, in order to check our results, the algorithm which performs the exact numerical evaluation of determinant has been implemented.

It has been previously reported, in the QCD case, that simulations with Lüsher technique suffer from the long autocorrelation [4] times. The source of these correlations was also suggested there. It is however important to learn whether they can be reduced by the selective refinement of the generation of the $\phi$ fields or while an update of the $A$ field. Below we discuss some improvements.

Tab.1 gives the results for the two observables: the average density of electrons $< n_\uparrow >$ and that of pairs of electrons with the opposite spin $< n_\uparrow(x) n_\downarrow(x) >$. The simple algorithm described above gives large autocorrelation times already for $K = 1$ and $U = 1$. Partly they could be caused by the critical slowing down introduced by the $\phi_k$ fields, especially those with small $\beta_k$. Since the model is in fact gaussian in $\phi$ fields, it is natural to apply multigrid (MG) methods [6]. Indeed the MG generation of $\phi$ fields reduces the autocorrelation times substantially (cf. row 3). However it introduces additional CPU coast which is practically prohibitive on larger lattices. The maximal decorrelation of the auxiliary fields is achieved by the independent generation of the eigenmodes of the quadratic form, see row 4. This requires the inversion of the N matrices and the results are comparable to the MG case with W-cycle.

Neither simple version nor MG refinement is capable to reproduce the exact determinant results for $U = 2$ (cf. column 3 of Table 1). The system did not thermalize even after $\sim 20$ times more thermalization steps then required for $U = 1$. This fact can be simply understood.

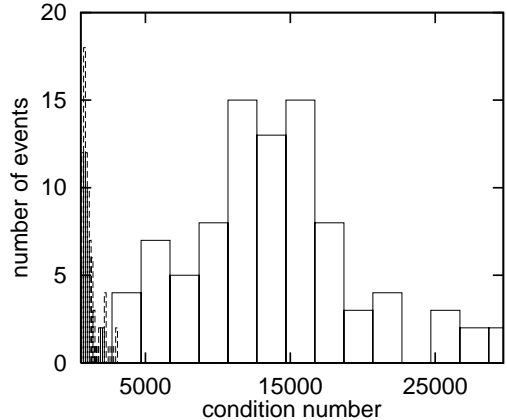

Figure 1. The distribution of the conditioning numbers for the $\mathcal{M}^\dagger\mathcal{M}$ matrix (solid line) and preconditioned matrix $\mathcal{M}^\dagger D^{-1}\mathcal{M}$ (dashed line)

When $U$ increases the spectrum of matrix $\mathcal{M}^\dagger\mathcal{M}$ becomes exponentially wider and the bigger number of $\phi$ fields in polynomial approximation is needed. The constraints imposed by large number of $\phi$ fields on one $A$ field becomes more restrictive and the mobility of algorithm rapidly decreases. Performing updates of $A$ fields more frequently is only a partial solution.

More promising is the direct preconditioning of the bosonic matrix. We define the preconditioning procedure as follows. Let $D$ be the diagonal part of $\mathcal{M} = A + D$. Then the matrix $\mathcal{M}^\dagger D^{-1}\mathcal{M}$ is better conditioned as can be seen from Fig.1. This effect is also clearly visible in the last row of Table 1. Our preconditioning reduces the autocorrelation time by more than a factor of 8. As a consequence wider range of couplings and lattice sizes becomes available. For largest lattice correlation time had large errors due to the shorter (relative to the $\tau_{int}$ measurement time). As the conditioning number of $\mathcal{M}^\dagger\mathcal{M}$ is largely governed by factor $\sqrt{\beta U/N_t}$ one can reduce the conditioning numbers by choosing large enough $N_t$. Our recent simulations performed on the $6^2 14$ lattice confirm this expectations. However still a lot of work remains to be done in order to reach

Table 1

|  | $\beta = 1, U = 1$ lattice $5^3$ | $\beta = 1, U = 1$ lattice $6^2 8$ | $\beta = 1, U = 2$ lattice $6^2 8$ | $\beta = 1, U = 1$ lattice $6^2 14$ |
|---|---|---|---|---|
| Exact determinant | 0.460(2) | 0.473(2) | 0.462(4) | — |
|  | 0.2197(2) | 0.2203(4) | 0.195(1) |  |
|  | $\tau_{int} = 2$ | $\tau_{int} = 2$ | $\tau_{int} = 3$ |  |
| Simple program | 0.461(5) | 0.471(6) | — | — |
|  | 0.2180(8) | 0.221(1) |  |  |
|  | $\tau_{int} = 570$ | $\tau_{int} = 870$ |  |  |
| Simple program with MG (W-cycle) | — | 0.470(4) | — | — |
|  |  | 0.2194(6) |  |  |
|  |  | $\tau_{int} = 160$ |  |  |
| Global generation of gaussian fields | 0.449(7) | — | — | — |
|  | 0.221(1) |  |  |  |
|  | $\tau_{int} = 170$ |  |  |  |
| Simple program with preconditioning | — | 0.470(5) | 0.463(7) | 0.475(7) |
|  |  | 0.220(1) | 0.195(1) | 0.186(3) |
|  |  | $\tau_{int} = 100$ | $\tau_{int} = 200$ | $\tau_{int} = 600$ |

Results for the density of electrons (upper) and those for the density of pairs (lower). The integrated autocorrelation times $\tau_{int}$ obtained from both observables are similar. They are quoted in units of the single update (sweep) of the $A$ field.

the physically interesting values of the couplings while keeping $\mathcal{M}^\dagger \mathcal{M}$ reasonably well conditioned $\beta U / N_t \simeq 0.5$ [7].

The recent modification of Lüscher method [8] is very promising. Additional global Metropolis step, which makes the approach exact, allows also better control of the number of auxiliary fields.

In conclusion, the Hubbard Model is in the Lüscher class, i.e. it can be mapped onto a system of bosons with local interactions. The half filled case has the positive Boltzman factor which admits standard Monte Carlo techniques. Preconditioning of the fermionic matrix strongly reduces the large autocorrelation times. Further studies are needed to turn this approach into viable alternative to the existing methods.

J. W. thanks B. Bunk and Ph. de Forcrand for the discussion.